\documentclass[twocolumn,showpacs,preprintnumbers,amsmath,amssymb]{revtex4}

\usepackage{graphicx}
\usepackage{dcolumn}
\usepackage{bm}

\def\be{\begin{equation}}
\def\ee{\end{equation}}
\def\ba{\begin{eqnarray}}
\def\ea{\end{eqnarray}}

\newcommand{\bea}{\begin{eqnarray}}
\newcommand{\eea}{\end{eqnarray}}

\def\nn{\nonumber}
\def\calS{{\cal S}}

\begin{document}

\preprint{hep-th/0611273; DAMTP-2006-116;
SPHT-T06/156; UB-ECM-PF-06-40}

\title{Ultraviolet properties of Maximal Supergravity}
\author{Michael B. Green}
\affiliation{ Department of Applied Mathematics and
Theoretical Physics\\
Wilberforce Road, Cambridge CB3 0WA, UK}
\author{Jorge G. Russo}
\affiliation{Instituci\' o Catalana de Recerca i Estudis Avan\c{c}ats (ICREA),\\
University of Barcelona, Av.Diagonal 647,  Barcelona 08028 SPAIN}
\author{Pierre Vanhove}
\affiliation{Service de Physique Th\'eorique, CEA/Saclay, F-91191Gif-sur-Yvette, France}

\date{\today}

\begin{abstract}
We argue that recent results in string perturbation theory indicate
that the four-graviton amplitude of four-dimensional $N=8$ supergravity
might be ultraviolet finite up to eight loops.
We similarly argue that the $h$-loop
$M$-graviton amplitude might be finite for $h<7+M/2$.

\end{abstract}

\pacs{11.25.-w,04.65.+e,11.25.Db}

\maketitle

Maximally extended supergravity has for a long time held a privileged position among supersymmetric field theories.
Its four-dimensional incarnation as $N=8$ supergravity \cite{n=8} initially raised the hope of a
perturbatively finite quantum theory of gravity while its origin in $N=1$ eleven-dimensional supergravity
\cite{Cremmer:1978km} provided the
impetus for the subsequent development of M-theory, or the non-perturbative completion of string theory.  Maximal
supergravities in various dimensions arise as special limits within type II string theory, which is free of ultraviolet
divergences.  However, the fact that higher-dimensional maximal
supergravity
is not
renormalizable means that it cannot be quantized in any conventional manner in
isolation from string theory.

Another well-known problem with theories such as type II supergravity, is that it is
notoriously difficult to analyze the constraints implied by maximal supersymmetry in a
fully covariant manner since there is no practical off-shell formalism that makes manifest the full supersymmetry.
For example, it has not yet been possible to determine the extent to which supersymmetry protects operators against
perturbative corrections.   However, a number of indirect arguments suggest that such protection may be far greater
than earlier estimates might suggest. For example, a certain amount of
evidence has accumulated over the past few years that the sum of the  Feynman diagrams
 of maximal supergravity
at a given number of loops may be less ultraviolet divergent than
expected \cite{Bern:1998ug}.  This is largely based on uncovering
fascinating connections with the diagrams of $N=4$ Yang--Mills, which are known to be ultraviolet
finite in four dimensions.  A different approach is to study
implications of string/M-theory duality
for the scattering amplitudes of type II supergravity.  In \cite{Green:2006gt}
we considered the $L$-loop Feynman diagrams of
the four-graviton scattering amplitude in eleven-dimensional
supergravity compactified on a two-torus and its string theory interpretation.  The lack of information
about the short-distance structure of M-theory is reflected by the nonrenormalizability of supergravity and this
ignorance was parameterized by an unlimited number of unknown coefficients of counterterms.
Nevertheless, requiring the structure
of the amplitude to be consistent with string theory led to interesting constraints.
Among these were strong nonrenormalization conditions in the
ten-dimensional type IIA string theory limit -- where the eleven-dimensional theory is
compactified on a circle of radius $R_{11}$.  This condition followed from dimensional analysis
together with the fact that the
string coupling constant is given by $e^{\phi} = R_{11}^{3/2}$,
where $\phi$ is the dilaton.  These conditions imply that the
genus-$h$ four-graviton amplitude has a low-energy limit that begins
with a power of $\calS^{h}\, R^4$, where $\calS^h$ is a symmetric monomial of
power $h$ made out of the Mandelstam invariants, $s$, $t$ and $u$.
This was interpreted as an indication that the
$h$-loop amplitude obtained in the low energy supergravity limit has
milder ultraviolet behaviour than naive expectations and four-dimensional
$N=8$ supergravity might
be free of ultraviolet divergences.

However, the arguments of \cite{Green:2006gt} were rather indirect.  Here we will proceed more directly and more
conservatively by using perturbative string theory as a regulator of the
ultraviolet divergences of supergravity.   We will see that the nonrenormalization conditions
of perturbative type II string theory obtained by Berkovits \cite{Berkovits:2006vc}, which are weaker than those proposed in
\cite{Green:2006gt}, point to the possible absence of ultraviolet
 divergences in the four-graviton amplitude of four-dimensional
$N=8$ supergravity amplitude up to eight loops.

We begin by noting that the  $h$-loop contribution to the four-graviton amplitude in ten-dimensional string theory has
the form
\be
A_4^h = {\alpha'}^{\beta_h-1}\, e^{2(h-1)\phi}\,\sum_i \calS_i^{(\beta_h)} \,I^{(h)}_i (\alpha' s,\alpha' t,\alpha' u)\, R^4\, ,
\label{tendimstring}
\ee
where $R$ is the Weyl curvature
and $\calS_i^{(\beta_h)}$ are monomials of power
$\beta_h$ in the Mandelstam invariants,
$s$, $t$ and $u$.
 Explicit one-loop \cite{Green:1981yb} and two-loop \cite{Bern:1998ug}
calculations show that $\beta_1=0$ and $\beta_2=2$.
In fact, recent perturbative superstring
calculations \cite{Berkovits:2006vc} determine that  $\beta_h = h$ up to five loops ($h=5$).
The less direct arguments of \cite{Green:2006gt} make use of string/M-theory dualities to argue that $\beta_h = h$
might hold to all orders.   In the following we will not specify the value of $\beta_h$ until
we need to.
The functions $I^{(h)}_i$ are given by integrals over the
moduli space of the  $h$-loop string world-sheet.
The number of such terms depends on the genus.  For example, at two loops ($h=2$) there are three terms,
\ba
\sum_{i=1}^3 &&\calS_i^{(2)} \,
I^{(2)}_i(\alpha' s,\alpha' t,\alpha' u) =
s^2 I^{(2)}_1(\alpha' s,\alpha' t,\alpha' u)
 \nn\\
&&+ t^2 I^{(2)}_2(\alpha' s,\alpha' t,\alpha' u) +
u^2 I^{(2)}_3(\alpha' s,\alpha' t,\alpha' u)\,.
\label{gentwores}
\ea
The detailed evaluation of the functions $I^{(h)}_i$ for $h>2$ is a daunting task but here we will only be
concerned with the general structure of the amplitude.

We wish to consider the low-energy field theory limit  of $A_4^h$  obtained by expanding the scalar functions,
$I_i^{(h)}$, in the
limit $\alpha'\to 0$ while holding the ten-dimensional Newton coupling, $\kappa_{(10)}^2 = {\alpha'}^4\, e^{2\phi}$, fixed.
Since $I_i^{(h)}$, which is an integral over world-sheet moduli,
does not have poles in $s$, $t$, $u$ \cite{Berkovits:2006vc}
its low-energy expansion starts with a constant or logarithmic term.
At low energies the amplitude (\ref{tendimstring}) therefore takes
the symbolic form
\be
A_4^h \sim \kappa_{(10)}^{2(h-1)}\,
{\alpha'}^{3-4h+\beta_h}\, \calS^{(\beta_h)} \,(1 + O(\alpha' s))\, R^4\,,
\label{loopexpan}
\ee
where we have
not kept track of possible factors that are logarithmic in the Mandelstam invariants but which are, in principle,
defined precisely by evaluating the amplitude.
 The expression (\ref{tendimstring}) is finite in ten-dimensional string theory due to the
presence of the string length that provides an ultraviolet cutoff.  Indeed, the divergence of the
expression (\ref{loopexpan}) in the low-energy limit, $\alpha' \to 0$, translates into the ultraviolet divergence
of the sum of all the contributions to the supergravity $h$-loop amplitude.
After interpreting the inverse string length as an ultraviolet momentum cutoff,  $\Lambda \sim\sqrt{{\alpha'}^{-1}}$,
 (\ref{loopexpan}) becomes
\be
A_4^h \sim \kappa_{(10)}^{2(h-1)}\, \Lambda^{8h -6-2\beta_h}\,\calS^{(\beta_h)} \,(1 + O(\alpha' s))\, R^4\,.
\label{loopexpanft}
\ee
So the presence of the prefactor $\calS^{(\beta_h)}\, R^4$ means that
 the leading divergences,  $\Lambda^{8h+2}$, of individual $h$-loop Feynman diagrams
cancel and the ultraviolet divergence of the sum of diagrams is reduced by a factor of
$\Lambda^{-8-2\beta_h}$.

We are interested in the maximal supergravity limit in lower dimensions so we will consider compactifying
the string loop amplitude on a
$(10-d)$-torus (with the external momenta and polarizations oriented in the $d$ non-compact directions).
Now consider the low energy limit $\alpha'\to 0$ with the radii of the torus proportional to $\sqrt{\alpha'}$,
so that all the massive Kaluza--Klein states, winding-number states and excited string states decouple.
This leads to an expression for the $h$-loop supergravity amplitude in $d$ dimensions with cut-off $\Lambda_d$.
For example, for a square torus with all radii equal to $r\,\sqrt{\alpha'}$,
the expression for $A_4^h$ behaves as  $\kappa_{(d)}^{2(h-1)}\,\Lambda_d^{(d-2)h-2\beta_h -6}$,
where the $d$-dimensional Newton constant, given by $\kappa_{(d)}^2 = {\alpha'}^{(d-2)/2}\,  e^{2 \phi}\, r^{d-10}$,
is held fixed.  Consider the low energy limit in a dimension $d$ for which the power of $\Lambda_d$ is
positive, so that $(d-2)h > 2\beta_h + 6 >0$.  We will make a "smoothness assumption" that this power of $\Lambda_d$
does not increase in the process of taking the low energy
limit. Although this sounds plausible, it remains unproven and would fail if the leading low-energy
behaviour of the function $I^{h}_i(s,t,u)$ in the compactified theory were an inverse power
of $s$, $t$ or $u$.
With this assumption it follows that ultraviolet divergences are absent in dimensions for which
\be
d < 2 + {2\beta_h+6\over h}\, .
\label{dimdiv}
\ee
In dimensions that satisfy this bound the expression (\ref{loopexpanft}) is ill-defined since it contains a negative
power of the cut-off, which vanishes, whereas finite and infrared divergent terms that are non-vanishing are not
exhibited.  In this case the negative
power of $\Lambda_d$ is replaced by a negative power of $s$, $t$ and $u$ of dimension $[s]^{(d-2)h/2- \beta_h -3}$,
together with possible
logarithmic factors.  Infrared divergences arise for dimensions $d\le 4$ and presumably sum up in the usual fashion
to cancel the divergences due to multiple soft graviton emission \cite{Weinberg:1965nx}.
These features of the field theory limit are seen explicitly in the compactified one-loop
($h=1$) amplitude, which has  $\beta_1=0$ \cite{Green:1982sw}.  In that case the amplitude has
the form of a prefactor of $R^4$ multiplying a  $\varphi^3$ scalar field theory box diagram.
This is ultraviolet divergent when $d\ge 8$, and finite for $4< d <8$. The low-energy limit of
the two-loop ($h=2$) string theory expression reduces to the supergravity
two-loop amplitude which was considered in detail in
\cite{Bern:1998ug,Smirnov:1999gc}.
There it was shown that a power of $s^2$, $t^2$ or $u^2$ factors out of the sum of all supergravity
Feynman diagrams, so that $\beta_2=2$.  In this case the sum of Feynman diagrams has the form of
prefactor  $s^2\, R^4$ multiplying the sum of planar and non-planar $s$-channel
double box diagrams of  $\varphi^3$ scalar field theory, together
with corresponding $t$-channel and $u$-channel terms.
More recently the fact that $\beta_2=2$ has been confirmed by explicit
two-loop calculations in string theory \cite{twostringloops}.
In this case
ultraviolet divergences arise when $d\ge 7$ and the amplitude is finite for $4< d < 7$.

The value of $\beta_h$ for $h>2$ has not been established from
direct supergravity calculations beyond two loops, but it is strongly suspected that
 $\beta_h\ge 2$ for $h>2$.  Furthermore, in contrast to the $h=1$ and $h=2$ cases,
  the sum of Feynman diagrams for $h>2$ is unlikely to reduce to a prefactor
multiplying diagrams of $\varphi^3$ scalar field theory. That would require $\beta_h =2(h-1)$,
which would lead to finiteness in $d<6$ dimensions (for $h=3$ it would
also contradict the presence of a three-loop term in  $D^6R^4$ found in \cite{gv:D6R4}).
 Using the value $\beta_h =2$ (the least possible value)
 in (\ref{dimdiv}) leads to the absence of ultraviolet divergences when
 \be
d < 2 + 10/h\, ,\qquad\qquad h>1\, ,
\label{tenlim}
\ee
which appeared in \cite{Bern:1998ug}.  This shows that the first ultraviolet divergence in four dimensions
cannot arise until at least five loops.

The full extent to which the four-graviton amplitude is protected from ultraviolet divergences should become
clearer with a more complete understanding of the constraints implied by maximal supersymmetry.
These are difficult to establish in the absence of an off-shell supersymmetric
formalism. In theories with less supersymmetry
such protection is typically afforded to F-terms, which can be expressed as integrals over a subspace of
the complete superspace.  One might estimate the extent to which the derivative expansion of the four-graviton
amplitude is protected by using on-shell superfield arguments.
This is explicit in the pure spinor formalism of the superstring developed by
Berkovits \cite{Berkovits:2006vc}, in which terms of the form
$\calS^{(k)}\, R^4$ are F-terms for $k\le 5$ and get vanishing contributions from $h > k$ loops.
Such terms arise from integration over a subset of the 32 components of the
left-moving and right-moving Grassmann spinor world-sheet coordinates, $\theta_L$ and $\theta_R$.
As a result, the low energy limits of both type IIA and IIB superstring theories at $h$ loops have $\beta_h =h$
for $h=2,3,4,5$ and $\beta_h \ge 6$ for $h\ge 6$.
Since $\calS^{(h)}\, R^4$ terms are protected for $h\le 5$ the
$\calS^{(6)} \,R^4$ interaction can only arise for $h\ge 6$.  If we once more make the assumption that the power
of $\Lambda_d$ does not increase in the process of taking the low energy limit in $d$
dimensions, we can see from (\ref{dimdiv})
that ultraviolet divergences are absent   for the following cases
\begin{eqnarray}
d &<& 2 +18/h\, , \qquad\qquad h>5
\label{berklim}\\
d &<& 4 + 6/h \qquad\qquad h=2,\dots,5\, .
\label{goodlim}
\end{eqnarray}
This indicates that the nonrenormalization conditions in
type II string theory \cite{Berkovits:2006vc} lead to the ultraviolet finiteness of the four-dimensional
$N=8$ supergravity four-graviton amplitude up to at least eight loops ($h=8$).
Note that although type IIA and IIB
four-graviton  string theory amplitudes are equal only up to four loops ($h=4$) in
ten dimensions  \cite{Berkovits:2006vc},  they are identical at all
loops in the $d$-dimensional low-energy supergravity limit.

One of the benefits of the superfield description of F-terms is that it includes all
terms related by supersymmetry.  For example, interactions
involving higher powers of the curvature tensor of the form $\calS^{(k)}\, R^M$, are F-terms
if $2k + M <16$ \cite{Berkovits:2006vc} (where $\calS^{(k)}$
 is a monomial made of Mandelstam invariants of the $M$-particle amplitude).  In this case
 an extension of the fermionic mode counting in \cite{Berkovits:2006vc} shows that there are no  corrections
 beyond $h=k+M-4$ loops. Generalizing our earlier
analysis, this means that the $h$-loop $M$-graviton amplitude
 behaves as $\calS^{(h+4-M)}\, R^M$ for $M\le h+4$.
It follows that the cut-off dependence of this amplitude is
$\Lambda_d^{(d-4)h -6}$ for $h<4+M/2$ and
$\Lambda_d^{(d-2)h-14-M}$ for $h\geq 4+M/2$. Given the previous
analysis, this would  imply that
the $M$-graviton amplitude is finite in four dimensions ($d=4$) if $h<7+M/2$.
It is notable that these arguments suggest the absence of
 divergences that might have arisen according to various superspace
 arguments  \cite{Howe:1980th,Howe:1983sr,Howe:1988qz,Howe:2002ui}.

Finally, we return to the suggestion  \cite{Green:2006gt}
that $\beta_h = h$. This was motivated by an indirect argument based on
considerations of M-theory duality rather than direct string calculations and is therefore less
well established.  In particular, it is
not yet apparent how this condition can be motivated by supersymmetry.  In this case  there is an extra
power of $s,t$ or $u$ for every additional loop and the divergence of the $h$-loop integral is markedly
reduced.  Substituting $\beta_h = h$ in (\ref{dimdiv}), and making the
earlier smoothness assumption, it  follows
that ultraviolet divergences are absent when
\be
d < 4 + 6/h \qquad\qquad h\ge 2\, .
\label{bestlim}
\ee
If correct, this would imply that ultraviolet divergences are absent to all orders in the four-graviton amplitude of
four-dimensional maximal supergravity. Finiteness of the four-graviton amplitude
suggests finiteness of all $M$-point functions since they are interconnected by unitarity.
Indeed, the arguments in \cite{Green:2006gt} have an obvious
extension to
multi-graviton amplitudes, which suggests that the $\calS^{(k)}\, R^M$ interactions
again have a dependence on the cut-off of the form $\Lambda_d^{(d-4)h -6}$.  This leads to the same condition,
(\ref{bestlim}), for ultraviolet finiteness of $M$-graviton amplitudes as in the four-graviton case.

The bound (\ref{bestlim}) is the same as the condition for the absence of
ultraviolet divergences in maximally supersymmetric  Yang--Mills
theory, which is known to be finite in four dimensions.
Indeed, the work of \cite{Bern:1998ug} points to
connections between loop amplitudes of maximal supergravity and those of maximal super-Yang--Mills motivated in part
by the  Kawai--Lewellyn--Tye relations \cite{Kawai:1985xq} that connect tree-level open and closed
string theory.  This suggests that
$N=8$ supergravity may be more finite than previously expected \cite{Bern:2005bb,Bern:2006kd}.

To summarize, in this paper we have considered the implications for
low energy $d$-dimensional
supergravity of the recently  discovered nonrenormalization properties of higher-genus contributions
to the four-graviton amplitude in type II superstring theory
\cite{Berkovits:2006vc}.  The relevant limit involves
compactification of string theory on a $(10-d)$-torus followed by a low
energy expansion at fixed Newton constant, $\kappa^2_d$.  We
argued that, subject to an important smoothness assumption, the
four-graviton amplitude of $N=8$ supergravity has no ultraviolet
divergences up to at least eight loops.   The nonrenormalization
conditions were obtained in \cite{Berkovits:2006vc} by using the pure spinor
formulation of string perturbation theory, in which
$\calS^{(h)}\, R^4$ is an F-term and gets no corrections beyond $h$
loops if $h\le 5$.  Similarly, we also argued that
the $M$-graviton amplitude is ultraviolet finite when $h< 7+M/2$.

An important further issue that remains to be resolved is the fact that, in
the low energy limit under consideration, infinite towers of states in
the nonperturbative sector of string theory (wrapped $Dp$-branes,
Neveu--Schwarz branes,  Kaluza-Klein charges and Kaluza--Klein
monopoles) become massless.  These are likely to give singular
contributions that may cast doubt on the validity of the perturbative
approximation, which only takes into account the perturbative
states \footnote{Similar observations have independenty been made by Hirosi
  Ooguri and John Schwarz.}.

It would obviously be of interest if this understanding could be extended
 to derive the all-orders non-renormalization conditions proposed in \cite{Green:2006gt}.  Interestingly,
there are similar situations in highly supersymmetric theories in
which an infinite number of higher-dimension operators are protected from
renormalization even though a naive application of supersymmetry would suggest that only a finite number should be
 \cite{Sethi:1999qv,Dine:1999jq}.

A priori, finiteness of $N=8$ seems very unlikely and, if true,  would cry out for a natural
explanation.
One possible framework for such an explanation might be a variant of twistor string theory
\cite{Witten:2003nn}, which naturally describes $N=4$ Yang--Mills coupled to superconformal gravity
\cite{Berkovits:2004hg,Berkovits:2004jj}. Perhaps one of the proposals for a $N=8$ twistor string theory
in  \cite{Abou-Zeid:2006wu} is on the right track.

\medskip

\begin{acknowledgments}
We are grateful to Nathan Berkovits for many useful interactions and to Savdeep Sethi  and
Edward Witten for correspondence.  We are also grateful to Hirosi
Ooguri and John Schwarz for communications regarding the possible
r\^ole of massless `non-perturbative' states.
P.V. would like to thank the LPTHE of Jussieu for hospitality where part of this work was carried out.
J.R. also acknowledges support by MCYT FPA 2004-04582-C02-01.
This work was partially supported by the RTN contracts
MRTN-CT-2004-503369, MRTN-CT-2004-512194 and MRTN-CT-2004-005104  and by the ANR grant BLAN06-3-137168.
\end{acknowledgments}

\end{document}